\documentstyle[epsfig]{mn}

\begin{document}
\title[Central Lopsidedness due to off-centred halo]
{Off-centred dark matter halo leading to strong central disc lopsidedness}
\author[C. Prasad  and C.J. Jog ]
       {Chaitanya Prasad$^{1}$\thanks{E-mail : prasadchatty@gmail.com} and
        Chanda J. Jog$^{2}$\thanks{E-mail : cjjog@physics.iisc.ernet.in}\\
$^1$  Department of Physics, Indian Institute of Technology - Bombay, Mumbai 400076, India\\
$^2 $ Department of Physics,
Indian Institute of Science, Bangalore 560012, India \\
}

\maketitle

\begin{abstract}
There is increasing evidence now from simulations and observations that the centre of
dark matter halo in a Milky Way type galaxy could be off-centred by a few 100 pc
 w.r.t. the galactic disc. We study the effect of such an offset halo on the orbits and kinematics
in the central few kpc of the  disc.
The equations of motion in the disc plane can be written
in terms of the disc and halo potentials when these two are concentric and a perturbation
term due to the offset halo. This perturbation potential shows an $m=1$ azimuthal variation, or is lopsided,
and its magnitude increases at small radii.
On solving these equations,
we find that the perturbed orbit shows a large deviation $\sim 40 \%$ in radius at R=1.5 kpc, and also strong kinematical lopsidedness.
Thus even a small halo offset of 350 pc  can induce surprisingly strong spatial and kinematical lopsidedness
in the central region within $\sim 3$ kpc radius. Further, the disc would 
remain lopsided for several Gyr, as long as
the halo offset lasts. This would have important implications for  
the dynamical evolution of this region.
\end{abstract}
\section{Introduction}
It is well-known that a galactic disc is located within an extended dark 
matter halo, whose existence is deduced from a nearly flat rotation curve (Rubin 1983, Binney \& Tremaine 1987).
The dark matter halo dominates the galaxy mass and hence the disc dynamics in the
outer regions of a spiral galaxy. In the inner central region of $\sim 5$ kpc radius, however, the disc dominates the
dynamics as in the Milky Way (Sackett 1997, Klypin et al. 2002).
 In theoretical as well as observational work, the dark matter halo is generally assumed to be concentric w.r.t. the disc
for simplicity.

There is increasing evidence that the dark matter halo of a Milky Way type galaxy
may be located off-centred w.r.t. the disc centre,  as seen in the simulations (Kuhlen et al. 2013) 
where the typical separation is 300-400 pc. Once it is generated, the offset feature seems to be long-lived, lasting for 
several Gyr. There is also an observational evidence for the halo being off-centred
w.r.t. the disc in a field galaxy, M94 (Chemin et al. 2016), and for a galaxy in the galaxy cluster Abell 3827 (Massey et al. 2015).
The dynamical origin of this offset is not clear although a few ideas such as excitation by a stellar bar
have been suggested 
(Kuhlen et al. 2013). 
A lower resolution may give only an upper limit to the offset, for example Schaller et al. (2015) find that more than 95 \% of simulated galaxies in their sample have an offset between their stars and dark matter that is smaller than the gravitational softening length of 700 pc they use.
The dynamics of an offset halo in a disc could well have some similarities
with the unsettled behaviour seen in central regions of galaxies (Miller \& Smith 1992). Such sloshing or unrelaxed central region is seen in mergers as well as in isolated galaxies (Jog \& Maybhate 2006), though this phenomenon has received little attention.

Motivated by this, in this paper we study how the spatial distribution and kinematics in the central disc is affected
by the off-centred dark matter halo, in a region where the disc is the dominant mass component. 
The halo potential can be written in terms of its value w.r.t. the disc centre and a
small perturbation which  takes a lopsided form. Next, the equations of
motion in the disc plane are solved under this joint disc plus halo system.
We find that the  offset halo results in a surprisingly large disturbance  over the central
few kpc. This effect is strongly dependent on the radius.
For typical exponential disc and pseudo-isothermal halo parameters for the Milky Way, and assuming an
offset of 350 pc as seen by Kuhlen et al. (2013) in their simulations, the resulting radial variation in the orbit 
is $\sim 40 \%$ at R=1.5 kpc which drops down to $\sim 3 \%$ at 3 kpc. 
The kinematics is also strongly disturbed.
Thus even a small offset of 350 pc can induce significant spatial and kinematical lopsidedness
in the central $\leq 3$ kpc region.

Section 2 contains the formulation of equations and section 3 contains the results, 
Sections 4 and 5 contain the discussion and conclusion respectively. 

\section {Formulation of equations}
We use cylindrical co-ordinates (R, $\phi$, z). 
The disc is taken to have an exponential surface density distribution (Freeman 1970), and the 
dark matter halo is taken to have a screened or pseudo-isothermal spherical density profile (Binney \& Tremaine 1987).
\subsection  {Potential due to a disc and off-centred halo}
The potential at a point $P$ in the galactic disc plane due to the disc and the off-centred halo
can be written in terms of the disc and halo contributions w.r.t. the disc centre, plus
the additional perturbation
term which arises due to the offset between the disc and the halo centres. 
We show later that this additional term is indeed a perturbation. This follows from the fact that in a typical  spiral galaxy like our Galaxy, the disc 
dominates the halo dynamics in the central region.

The disc centre and the halo centre are denoted by $C_D$ and $C_H$, and the offset 
between the two is denoted by a distance $d$. The schematic diagram is given in Figure 1.
Note that $\phi$ is the angle between the radius vector to the point $P$ from the disc centre and the
direction between the disc centre and the halo centre. Note that the figure is not drawn to scale, where for the sake of clarity we have
shown $d \sim R$  whereas in our calculation we assume $d << R$.
\begin {figure} 
  \includegraphics[height=1.7in,width=2.3in]{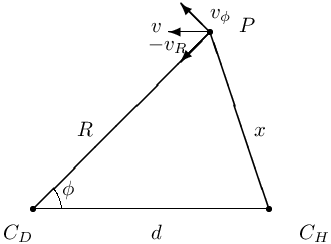}
   \caption{Schematic diagram showing the centres of disc and an off-centred halo, separated by a distance $d$, and their effect at point P in the disc plane}
 \end{figure}
The density distribution of an exponential disc and a pseudo-isothermal halo are given respectively 
as
$$    \Sigma (R) = \Sigma_0 exp(-\frac{R}{R_D}) \eqno (1)$$
and,
$$ \rho (r) = \frac{\rho_0}{1+\frac{r^2}{R_C^2}}  \eqno (2) $$
where $r$ is given in spherical co-ordinates. Here $R_D$ is the exponential disc scale-length and $R_c$ is the halo core radius.

Solving the Poisson equation, the potential in the mid-plane at a point $P$  due to the disc
is given by (see e.g., Binney \& Tremaine 1987):
$$    \Phi_D= -\pi G \Sigma_0 R \: [I_0(y)K_1(y)-I_1(y)K_0(y)] \eqno(3)$$
where, $ y={R}/{2R_D} $. 

The halo mid-plane potential at the point $P$ when the halo is concentric w.r.t. the disc   is obtained as:
$$    {\Phi_H} (d=0) = 4\pi G \rho_0 R_C^2 \: [\frac{\log(R_C^2+R^2)}{2} + \frac{R_C}{R}\arctan(\frac{R}{R_C}) - 1] \eqno(4) $$
The halo mid-plane potential at the point $P$ when the halo is off-centred is given by,
$$    \Phi_H = 4\pi G \rho_0 R_C^2[\frac{\log(R_C^2+x^2)}{2} + \frac{R_C}{x}\arctan(\frac{x}{R_C}) - 1] \eqno(5) $$
\noindent We consider small values of d ($<< R$), so that
$$    x^2 = R^2+d^2-2dR\cos\phi \\$$
$$    \approx R^2(1-\frac{2d\cos\phi}{R}) \eqno(6) $$
Simplifying each term separately, we get
$$    \log(R_C^2+x^2) =\log(R_C^2+R^2-2dR\cos\phi) \\ $$
$$    \approx \log(R_C^2+R^2) -\frac{2dR\cos\phi}{R_C^2+R^2} \eqno (7) $$
$$    \frac{R_C}{x}\arctan(\frac{x}{R_C}) =\frac{R_C}{R(1-\frac{2d\cos\phi}{R})^\frac{1}{2}}\arctan(\frac{x}{R_C}) \\ $$
$$    \approx \frac{R_C}{R}\arctan(\frac{R}{R_C})+\frac{R_C d\cos\phi}{R^2}\arctan(\frac{R}{R_C}) \eqno (8) $$
Therefore, the first terms from both expressions are simply terms from a halo that is concentric with the disc, and the other two terms arise from the non-zero value of d, with the disc potential remaining undisturbed. Thus, the total halo potential, $\Phi_H$ is given by,
$$\Phi_H = \Phi_H (d=0)     +  4\pi G \rho_0 R_C^2 d\cos\phi \: \times $$
$$   \quad \quad \quad \quad \quad \quad \quad \quad \quad \quad [\frac{R_C}{R^2}\arctan(\frac{R}{R_C})-\frac{R}{R_C^2+R^2}] \eqno (9) $$
where the second term on the r.h.s. defines the perturbation term $\Phi_{pert}$.
Thus, the total potential can be written as,
$$    \Phi=\Phi_D+\Phi_H(d=0)+\Phi_{pert} \eqno (10) $$
where the first two terms give the potential due to disc and halo with the same centre, $C_D$, and the last term is the perturbation potential due to the halo being offset w.r.t. the disc centre.

\subsection {Perturbation potential with a lopsided form}
\noindent The perturbation potential (see eq. 9) is thus given by
$$    \Phi_{pert} (R, \phi) = \Psi (R) \cos\phi \eqno(11) $$
where,
$$    \Psi(R) = 4\pi G \rho_0 R_C^2 d \: [\frac{R_C \arctan\frac{R}{R_C}}{R^2}-\frac{R}{R_C^2+R^2}] \eqno(12)$$
\noindent The perturbation potential depends linearly on 
cos $\phi$, thus it has a lopsided form corresponding to an $m=1$ azimuthal perturbation (e.g., Jog \& Combes 2009).
This is as expected from the geometry of the mass distribution (see the schematic plot in Fig. 1).
Its magnitude is linearly proportional to the separation $d$ between the disc and the halo centres.
Thus, as the offset $d \rightarrow 0$, the perturbation goes to zero and the net potential then is simply the sum
of the potential due to the disc and a concentric halo.

Next we define $\epsilon_{pert}$, the dimensionless perturbation potential, to be the ratio of the 
perturbation potential over the total undisturbed contributions of concentric disc and halo:
$$ \epsilon_{pert} (R) = \frac {\Psi (R)}{ (\Phi_D  + \Phi_H) }  \eqno(11) $$
Using $\Psi(R)$ as the perturbation potential, and following the procedure as in Jog (1997) and Jog (2000), we can next derive the perturbed radius and the azimuthal and radial velocity components for a particle  in a closed orbit of radius R$_0$.

The perturbation in R is thus  obtained as:
$$    \delta R = -(\frac{2\Psi}{R}+\frac{\partial\Psi}{\partial R})_R{}_0\frac{\cos\phi_0}{\kappa^2-\Omega_0^2} \eqno (12) $$
where $\kappa$ and $\Omega_0$ denote the epicyclic frequency and the angular speed respectively at $R_0$.
$$    \Rightarrow \delta R = -\frac{d \: \cos\phi_0}{\kappa^2-\Omega_0^2}\frac{4\pi G \rho_0 R_C^2}{R_C^2+R_0^2}(\frac{R_C^2}{R_0^2}+\frac{2R_0^2}{R_C^2+R_0^2}-3)  \eqno(13)$$
The perturbation in the azimuthal velocity $v_\phi$ is governed by:
$$    R_0\frac{d^2\delta\phi}{dt^2}+2\Omega_0\frac{d\delta R}{dt} = \frac{\Psi \sin\phi_0}{R_0} \eqno(14) $$
Integrating w.r.t. time, this gives:
$$  R_0\frac{d\delta\phi}{dt} = - 2\Omega_0 {\delta R} - \frac {\Psi (R) cos\phi_0}{\Omega_0 R_0} \eqno(15)$$
The net velocity along the azimuthal direction is given by:
$$    v_\phi=v_c+R_0\frac{d\delta\phi}{dt}+\Omega_0\delta R \eqno(16) $$
Hence the perturbation velocity along the azimuthal direction is:
$$    \delta v_\phi = R_0\frac{d\delta\phi}{dt}+\Omega_0\delta R \eqno(17) $$
On combining with equation (15), this reduces to:
$$     \delta v_\phi = -(\frac{\Psi\cos\phi_0}{\Omega_0 R_0}+\Omega_0\delta R) \eqno(18) $$
In full this is given as:
$$    \delta v_\phi  = -4\pi G \rho_0 R_C^2 d\frac{\cos\phi_0}{\Omega_0}(\frac{R_C\arctan(\frac{R_0}{R_C})}{R_0^3}-\frac{1}{R_0^2+R_C^2}) \\ $$
$$    +  4\pi G \rho_0 R_C^2 d\frac{\cos\phi_0 \Omega_0}{\kappa^2-\Omega_0^2}\frac{1}{R_0^2+R_C^2}(\frac{R_C^2}{R_0^2}+\frac{2R_0^2}{R_0^2+R_C^2}-3) \eqno(19) $$
The perturbed radial velocity is given by:
$$ v_R= \frac{dR}{dt} = \frac{d(R_0+\delta R)}{dt} \eqno(20) $$
Using equation (13), this is given as
$$ v_R= \frac{\sin\phi_0}{\kappa^2-\Omega_0^2}\frac{4\pi G\Omega_0 \rho_0 R_C^2 d}{R_C^2+R_0^2}(\frac{R_C^2}{R_0^2}+\frac{2R_0^2}{R_C^2+R_0^2}-3) \eqno(21) $$
While discussing the results in Section 3 we refer to the radius as $R$ for convenience (without the subscript $0$).
\section {Results}
\subsection {Input parameters}
The above results 
are general, and are valid for a typical spiral galaxy where the disc dominates the
dynamics in the central region. 
We now consider the input parameters for the Galaxy and study the
effect of the off-centred halo on the central disc region.
We take the central disc surface density, $\Sigma_0 = 640.9$ M$_{\odot}$ pc$^{-2}$, and the
exponential disc scale-length, $R_D$ = 3.2 kpc; and the halo central density $\rho_0$ = 0.035 $M_{\odot} pc^{-3}$ and the core radius $R_c = 5$ kpc; these values are taken from the Galaxy mass model by Mera et al. (1998).
\subsection{ Resulting lopsidedness}
First, we plot the value of the dimensionless perturbation potential, $\epsilon_{pert} (R)$, defined in equation (11)
for a choice of $d = 350$ pc, see Fig. 2. 
\begin {figure} 
  \includegraphics[height=2.3in,width=2.7in]{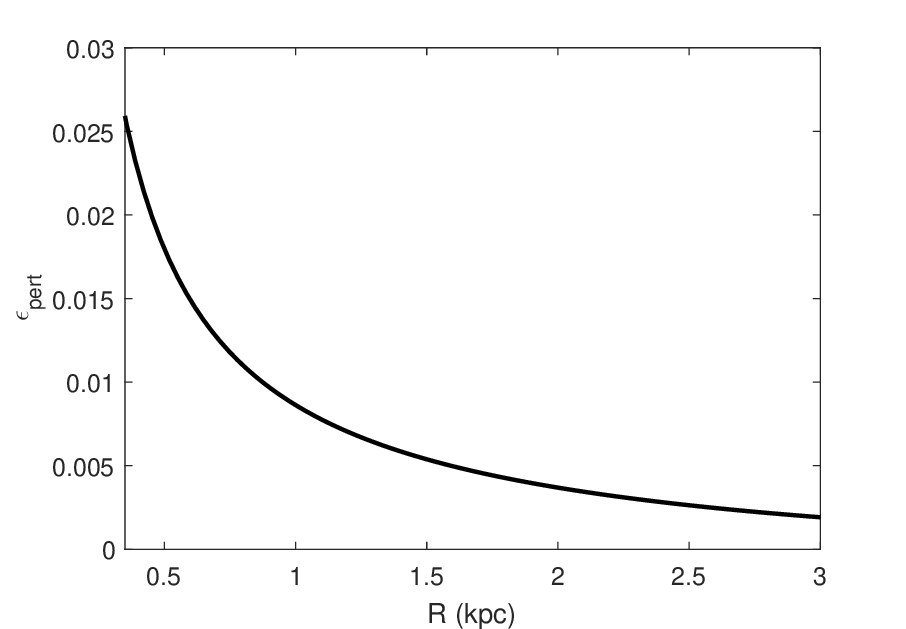}
   \caption{The perturbation potential $\epsilon_{pert}$ arising due to the halo that is off-centred by 350 pc w.r.t. the disc versus the radius R.  
}
 \end{figure}
We check that $\epsilon_{pert} (R)$ is indeed is a perturbation, justifying the assumption made in Section 2.1. The value is higher as $R$ decreases, this is expected physically, since the perturbative effect of halo would be stronger at a
smaller radius.
Note that the magnitude of the perturbation potential  
falls with increasing radius $R$. By $R = $ 1 kpc, the fractional value of the perturbation potential is less than 1 \%. 
Yet, as we will show next, this still results in strong spatial and  kinematical lopsidedness of the closed orbits in the disc, because
the perturbed equations have a strong dependence on radius.
\begin{figure}
\centering
\includegraphics[height=2.0in, width=2.4in]{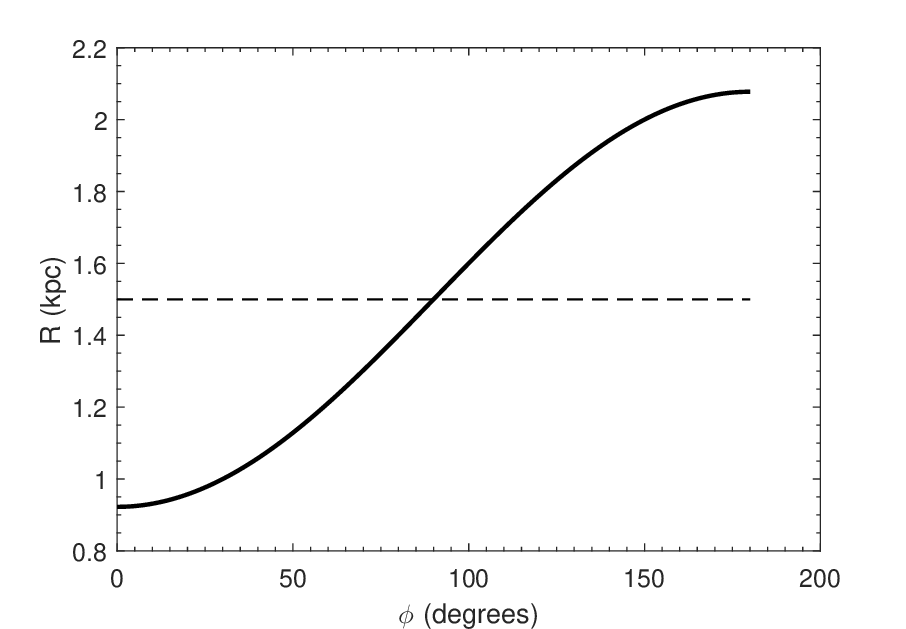}
\medskip 
\includegraphics[height=2.0in, width= 2.5in]{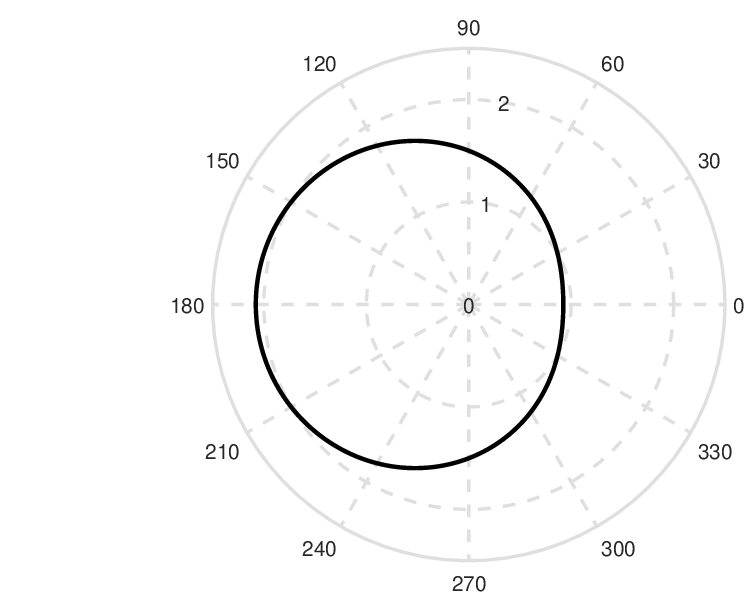} 
\bigskip
\includegraphics[height=2.0in, width=2.4in]{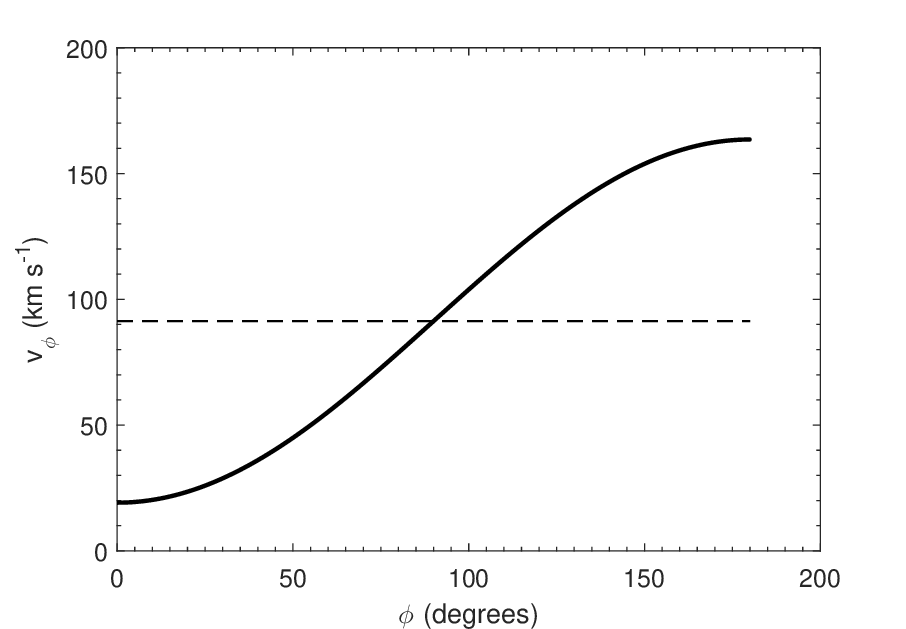}
\medskip
\includegraphics[height=2.0in, width=2.4in]{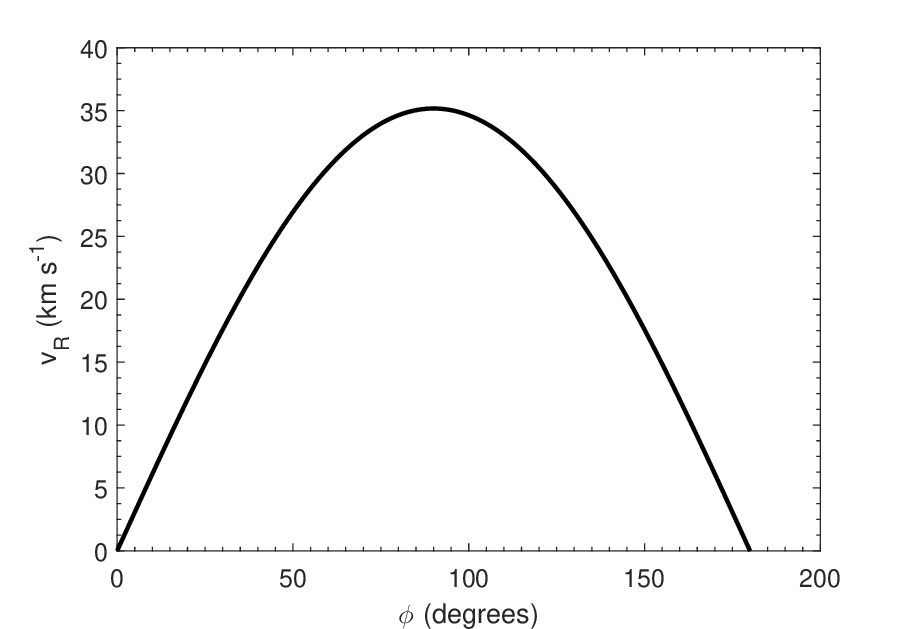}
\bigskip
\caption{Resulting orbit for an initial circular orbit at R=1.5 kpc, shown as R vs. $\phi$ in Fig. 3a, as a polar plot in Fig.3b, and the perturbed azimuthal and radial velocity components vs. $\phi$ are shown in Fig.3c and Fig.3d respectively. Figs. 3a and 3b show that the orbit is strongly distorted or lopsided spatially,
it is shortened along $\phi=0^0$ and elongated in the opposite direction, along $\phi=180^0$.} 
\end{figure}

 In Fig. 3a we plot the net radius of the orbit vs. $\phi$ 
for an initial circular orbit of radius, $R$ = 1.5 kpc. The values for the other half of the azimuthal range is not shown as it is a mirror image of this
since these quantities depend simply on $\phi$. Fig. 3b shows a polar plot of the perturbed orbit drawn against an unperturbed circle
showing the original circular orbit at R$_0$ = 1.5 kpc. Thus the perturbed orbit shows a typical lopsided form: it is shortened along $\phi=0^0$ by $\sim 40 \%$ and elongated along the opposite direction 
(i.e., along $\phi =180^0$).
The perturbed azimuthal and radial velocity components vs. $\phi$ are shown in Fig.3c and Fig.3d respectively. Note that the change in azimuthal velocity is huge: its maximum magnitude (which occurs at $\phi = 0^0$) is 72.2 km s$^{-1}$. The resulting radial or streaming velocity value is also significant =$35.2$ km s$^{-1}$, which occurs at $\phi = 90^0$. 
Thus, the radius and the velocity components of the perturbed orbit are strongly lopsided due to the perturbation potential resulting from the off-centred halo.

The resulting strong  lopsidedness is further illustrated in Fig. 4 where the azimuthal velocity due to concentric disc and halo are shown as well as the
perturbed value obtained at $\phi = 0^0$ as a function of $R$. The net unperturbed velocity is given by the square root of these two added in quadrature. The magnitude of the perturbed azimuthal velocity increases rapidly below 1.5 kpc, and
can be comparable to the original rotational velocity, and the velocity has a negative sign.
The high negative values of the perturbed azimuthal velocity indicates that the perturbed orbit is no longer a 
nearly circular orbit, rather it takes on the shape of a radial orbit. This has interesting dynamical implications for the central region, as discussed in Section 4.
 
\begin {figure} 
   \includegraphics[height=2.3in,width=2.7in]{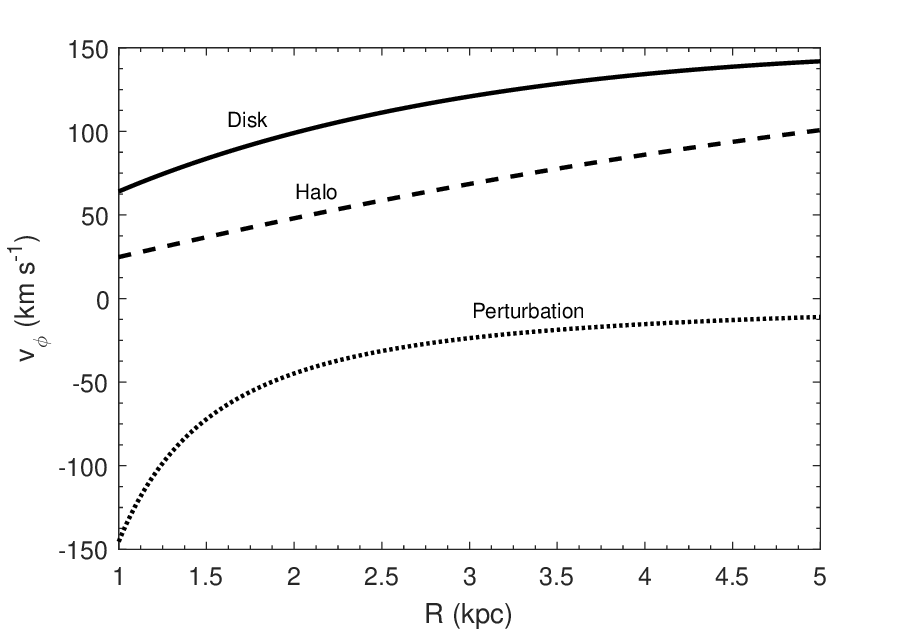}
   \caption{The azimuthal velocity components vs. the radius R for the disc-alone and the halo, both with the same centre;
and the perturbed azimuthal velocity component along $\phi = 0^0$, arising when the halo is taken to be offset w.r.t. the disc centre by 350 pc. 
Note that the disc dominates the dynamics in the central few kpc, and the perturbed velocity component is large.}
  \end{figure}
The fractional values of the change in the radius and the azimuthal velocity for the perturbed motion w.r.t. the undisturbed values are shown in Fig. 5a and Fig. 5b respectively. 
Clearly a significant radial range up to R = 3 kpc is strongly disturbed as seen from the spatial and kinematical lopsidedness.
In fact the assumption of linear perturbation in the calculation is not strictly valid 
 below R $ < 1.2 $ kpc. 
Note that at larger radii (beyond 3.2 kpc), the sign of the radial change is opposite so that it is positive at $\phi = 0^0$,
however, the magnitude of the fractional change is very small (see Fig. 5a).
\begin {figure} 
\includegraphics[height=2.0in,width=2.4in]{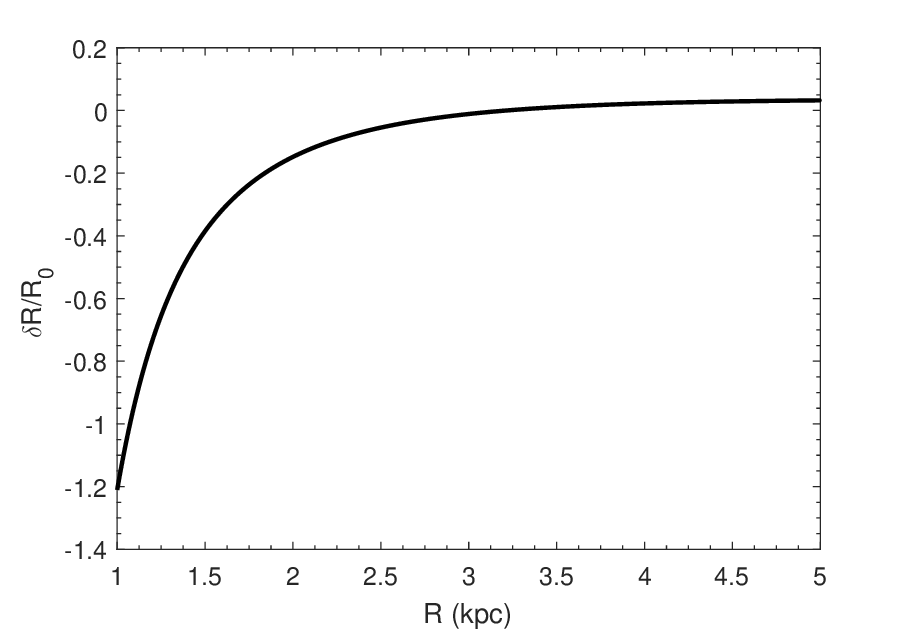}
\medskip 
\includegraphics[height=2.0in,width=2.4in]{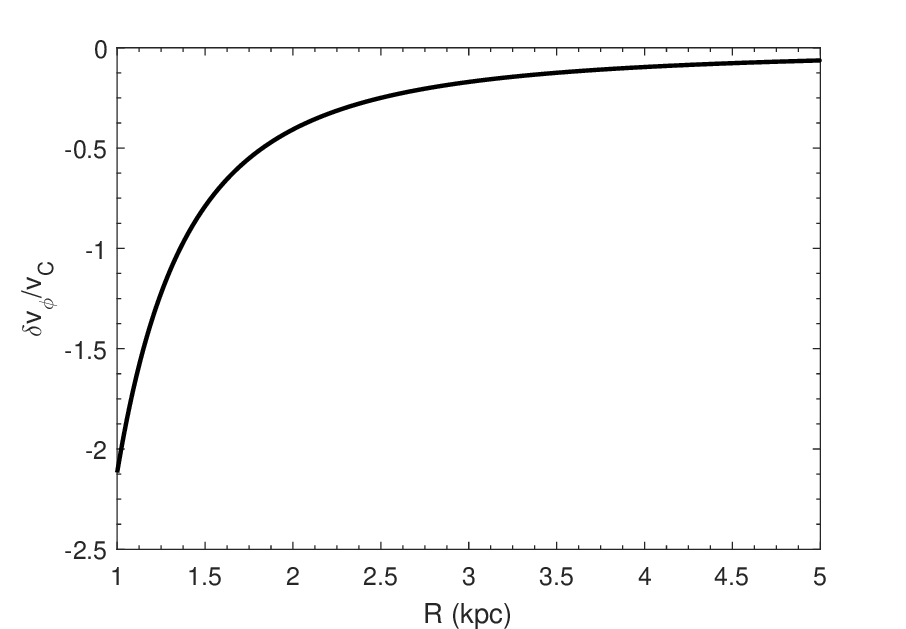} 
\bigskip
   \caption{Plot of the fractional change in the radius of the orbit (Fig. 4a, top panel) and fractional change in the azimuthal velocity component (Fig. 4b, lower panel)
both obtained at $\phi = 0^0$ vs. radius, R }
  \end{figure}

In Fig. 6 we show a polar plot of a range of circular orbits with initial radii, $R$ = 1.5, 2, 2.5 and 3 kpc as shown by grey
circles, and the corresponding perturbed orbits due to the off-centred halo - these are shown as a bold, dashed, larger dashed, and a bold orbit
respectively. This brings out in a striking way the strong azimuthal variation in the orbital radius, e.g. at $ R  = 1.5 $ kpc.

\begin {figure} 
  \includegraphics[height=2.3in,width=2.8in]{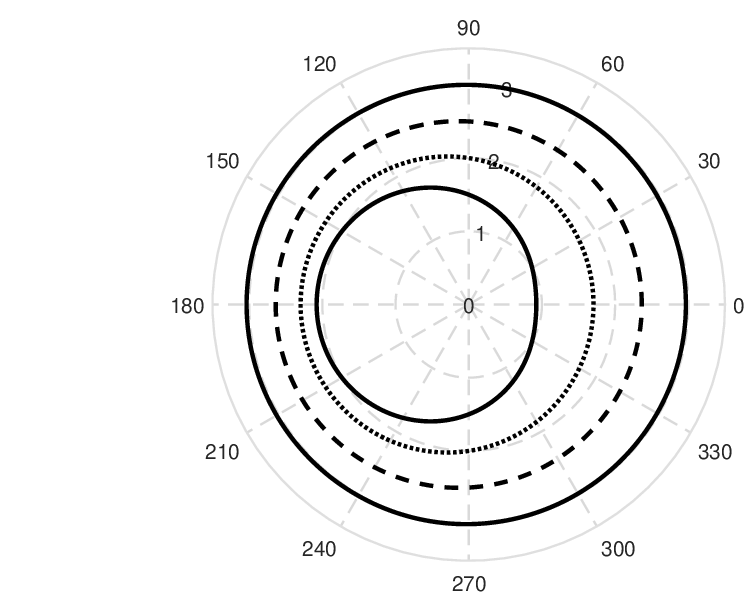}
   \caption{A polar plot of a range of circular orbits with initial radii, R= 1.5, 2, 2.5 and 3 kpc as shown by grey
circles, and the corresponding perturbed orbits due to the off-centred halo, shown as a bold, dashed, larger dashed, and a bold orbit.
Clearly the orbits within R=3 kpc are severely lopsided, especially at small R.
} 
 \end{figure}

As shown above, an off-centred dark matter halo leads to strong central disc lopsidedness. This
is in contrast to the fairly moderate lopsidedness seen in the outer regions of galaxies  
(Rix \& Zaritsky 1995, Jog 1997). The outer regions of galaxies are often observed to be lopsided where a radially constant perturbation potential gives a good fit to the observations (Rix \& Zaritsky 1995,  Angiras et al. 2006,  Zaritsky et al. 2013). In that case, at $\phi = 0^0$,
the typical fractional change in the radius and the azimuthal velocity  is respectively equal to twice the fractional perturbation potential or about 10\% (Jog 2000); and the fractional perturbation potential or 
about 5\% (Jog 2002). We note that, in contrast, for the off-centred halo considered here, the resulting perturbed spatial and velocity components have much higher values and show a strong radial dependence. This is because here the perturbed equations have a strong dependence on radius and  do not give a simple linear dependence on the 
dimensionless perturbation parameter $\epsilon_{pert}$.
\subsection {Parameter search}
We next check the dependence of the results on the input parameters. First we vary the halo concentration by taking $R_c = 4$ kpc and 6 kpc (instead of 5 kpc as was used above), these correspond respectively to a more centrally concentrated and a low mass halo and a more extended and massive halo. As expected, a lower core radius leads to a lower magnitude of the perturbation potential, 
$\epsilon_{pert}$.  
The resulting fractional variation in the orbital radius at R = 1.5 kpc is now  25\% for $R_c = 4$ kpc as opposed to 40\% obtained for $R_c = 5$ kpc, while a 
higher $R_c = 6 $ kpc results in a higher relative  fractional change in radius of 62\%.

Since the perturbation potential as well as the resulting radial and velocity changes are linearly proportional to the offset value $d$, a value of $d$ smaller than 350 pc that was used above will lead to correspondingly smaller perturbed values.
We stress that the origin as well as the saturation or a long-term existence of a such an offset is not understood, but presumably it 
is tied to the physical parameters of a Milky Way type galaxy. Thus we expect that the range 300-400 pc seen in the simulations by Kuhlen et al. (2013) is a genuine physical range, 
and we do not expect much variation around this range.
\section {Discussion}
\noindent 1. We have assumed that the disc and the halo do not move under the common gravitational field of each other.
This is justified from the constant, long-lived offset seen in the
simulations by Kuhlen et al. (2013). 

\noindent 2. 
The present paper  shows that in a typical spiral galaxy like the Galaxy, the dark matter halo that is offset by 350 pc w.r.t. the galactic disc centre
is sufficient to disturb the orbits and kinematics substantially within the central few kpc of the disc.
Since the offset is known to be long-lived, such resulting disc lopsidedness  will also be long-lived. 
This can have far-reaching implications:

First, the  resulting lopsided orbits 
may be important in feeding the AGN (active galactic nucleus) at the centre.
The highly disturbed, lopsided orbits in the central region may be responsible for causing an outward angular momentum transport,
and thus helping gas infall, similar to what was shown for lopsidedness in the outer disc (Saha \& Jog 2014), and in
agreement with the infall proposed for the central region (Combes 2016).
This could be important in fuelling a central AGN even in an isolated galaxy, which is a long-standing problem.

Second, 
the perturbed orbits with high azimuthal velocity dispersion  would
tend to be radial, especially for low $R$ (Section 3.2). This may make it easier to form a central triaxial structure or a  bar (Binney \& Tremaine 1987, 
Palmer \& Papaloizou 1987).
Kuhlen et al (2013) had suggested that a stellar bar causes the dark matter halo density to be off-centred 
and that a bar converts cusp to core via resonant interaction. However, one problem they noted with this picture is that
the bar strength varies with time.
If the orbits disturbed due to an off-centred halo could support bar formation, then this could help maintain the offset of the halo for a long time as seen in simulations. This qualitative idea should be checked by future high
 resolution simulations.
\section {Conclusion}
We study the effect of an off-centred  dark matter halo 
on the central galactic disc, where the disc dominates the dynamics. 
 The equations of motion are formulated in terms of 
disc and halo potential contributions when the two are concentric plus a perturbation term
that arises due to the halo offset. The perturbation potential has a lopsided form, that is, it shows an $m=1$ azimuthal dependence.
The magnitude of the perturbation potential depends on the halo parameters and also the offset distance $d$ between the disc and the halo centres.
 Though the fractional perturbation potential is small ($< 1 \%$)
we find that it has a surprisingly strong effect on the orbits and  the kinematics in the central region within a radius of about 3 kpc. 
The resulting strong lopsidedness can have important dynamical consequences, 
for example it can help fuel the central AGN via an outward angular momentum transfer.
An off-centred halo thus results in complex and interesting dynamics in the central region  which 
should be explored in future work.

\bigskip

\noindent {\bf Acknowledgments:} C.J. would like to thank the DST, Government of India for support via J.C. Bose fellowship (SB/S2/JCB-31/2014).

\bigskip

\noindent {\bf References}

\medskip

\noindent Angiras R. A.,  Jog, C.J.,  Dwarakanath, K.S.,  Verheijen, M.A.W.
 2007, MNRAS, 378, 276

\noindent Binney, J.,  Tremaine, S. 1987, Galactic Dynamics, Princeton: Princeton Univ. Press.

\noindent Chemin, L., Hure, J.-M., Soubiran, C., Zibetti, S., Charlot, S. , Kawata, D.
2016, A \& A, 588, 48

\noindent Combes, F. 2016, to appear in  IAU Symposium 322, The Multi-Messenger Astrophysics of the Galactic Centre, 
eds. S. Longmore, G. Bicknell and R. Crocker (arXiv:1608.04009)

\noindent Freeman, K.C. 1970, ApJ, 160, 811

\noindent Jog, C.J., \& Combes 2009, Phy Rep, 471, 75

\noindent Jog, C.J., \& Maybhate, A. 2006, MNRAS, 370, 891

\noindent Jog, C.J. 2002, A\&A, 391, 471

\noindent Jog, C. J. 2000, ApJ, 542, 216

\noindent Jog, C.J. 1997, ApJ, 488, 642

\noindent Klypin, A., Zhao, H., \& Somerville, R. S. 2002, ApJ, 573, 597

\noindent Kuhlen, M., Guedes, J., Pillepich, A., Madau, P., and Mayer, L. 
2013, ApJ, 765, 10

\noindent Massey, R. et al 2015, MNRAS, 449, 3393

\noindent Mera, D., Chabrier, G., Schaeffer, R. 1998, A\&A, 330, 953

\noindent Miller, R.H., \& Smith, B.F. 1992, ApJ, 393, 508

\noindent Palmer, P.L., Papaloizou, J. 1987, MNRAS, 224, 1043

\noindent Rix, H.-W., \& Zaritsky, D. 1995, ApJ, 447, 82

\noindent Rubin, V. 1983, Science, 220, 1339

\noindent Sackett, P.  1997, ApJ, 483, 103

\noindent Saha, K., \& Jog, C.J. 2014, MNRAS, 444, 352

\noindent Schaller, M., Robertson, A., Massey, R., Bower, R.G., Eke, V. R.
2015, MNRAS, 453, L58

\noindent Zaritsky, D. et al. 2013, ApJ, 772, 135

\end{document}